\documentclass[%
 twocolumn,
 superscriptaddress,
 preprintnumbers,
 nofootinbib,
 amsmath,amssymb,
 aps,
 prl,
]{revtex4}
\usepackage{psfrag,slashed,cancel,array,graphicx,hyperref}
\usepackage{subfigure}
\usepackage[utf8]{inputenc}
\usepackage{mathtools}
\usepackage{mathrsfs}
\usepackage{multirow}
\usepackage{braket}
\usepackage{booktabs} 
\usepackage[normalem]{ulem}
\hypersetup{pdftitle={},pdfcreator={},linkcolor=[rgb]{0.15,0.35,0.75},colorlinks=true,citecolor=[rgb]{0.675,0,0.2},urlcolor=[rgb]{0.15,0.35,0.65}}

\allowdisplaybreaks

\begin{document}


\title{Next-to-Next-to-Leading Order QCD Corrections to Polarized Semi-Inclusive Deep-Inelastic Scattering}

\author{Saurav Goyal}
\email{sauravg@imsc.res.in}
\affiliation{The Institute of Mathematical Sciences,  Taramani, 600113 Chennai, India}
\affiliation{Homi Bhabha National Institute, Training School Complex, Anushakti Nagar, Mumbai 400094, India}
\author{Roman N. Lee}
\email{r.n.lee@inp.nsk.su}
\affiliation{Budker Institute of Nuclear Physics, 630090, Novosibirsk, Russia}
\author{Sven-Olaf Moch}
\email{sven-olaf.moch@desy.de}
\affiliation{II. Institute for Theoretical Physics, Hamburg University, D-22761 Hamburg, Germany} 
\author{Vaibhav Pathak}
\email{vaibhavp@imsc.res.in}
\affiliation{The Institute of Mathematical Sciences, Taramani, 600113 Chennai, India}
\affiliation{Homi Bhabha National Institute, Training School Complex, Anushakti Nagar, Mumbai 400094, India}
\author{Narayan Rana}
\email{narayan.rana@niser.ac.in}
\affiliation{Homi Bhabha National Institute, Training School Complex, Anushakti Nagar, Mumbai 400094, India}
\affiliation{School of Physical Sciences, National Institute of Science Education and Research, 752050 Jatni, India}
\author{V. Ravindran}
\email{ravindra@imsc.res.in}
\affiliation{The Institute of Mathematical Sciences, Taramani, 600113 Chennai, India}
\affiliation{Homi Bhabha National Institute, Training School Complex, Anushakti Nagar, Mumbai 400094, India}

\date{\today}

\begin{abstract}
Polarized semi-inclusive deep-inelastic scattering (SIDIS) is a key process in the quest for a resolution of the proton spin puzzle.
We present the complete results for the polarized SIDIS process at next-to-next-to-leading order (NNLO) in perturbative quantum chromodynamics.
Our analytical results include all partonic channels for the scattering of polarized leptons off hadrons and a spin-averaged hadron identified in the final state.
A numerical analysis of the NNLO corrections illustrates their significance and the reduced residual scale dependence in the kinematic range probed by the future Electron-Ion-Collider EIC.
\end{abstract}
 


\maketitle

%
Deep-inelastic scattering (DIS) of leptons off hadrons provides
valuable information on the structure 
of hadrons at high energies in terms of their partonic constituents namely quarks, 
anti-quarks and gluons, and also of the underlying strong interaction dynamics through quantum chromodynamics (QCD) \cite{Blumlein:2012bf}. 
The DIS structure functions (SFs), encoding this information, are subject to QCD factorization that separates short-distance dynamics accessible in perturbation theory
from the long-distance (non-perturbative) one. 
The perturbative part, so-called coefficient functions (CFs), is 
computed in powers of the strong coupling $\alpha_s$, 
while the non-perturbative parton dynamics inside the hadron 
are parameterised in terms of parton distribution functions (PDFs), 
generally extracted from cross section data~\cite{Workman:2022ynf}.
Semi-inclusive DIS (SIDIS) with an identified hadron in the final state adds to the factorization formalism parton fragmentation functions (FFs)~\cite{Metz:2016swz},
which encode the parton dynamics in their recombination to
form hadrons.

Polarized DIS is a key process for the resolution of the long-standing proton spin puzzle. It gives access to the longitudinal spin structure of hadrons~\cite{Aidala:2012mv}, parameterised by helicity (spin-dependent) PDFs~\cite{deFlorian:2009vb,deFlorian:2014yva,Nocera:2014gqa}. 
The proton spin can be determined from a sum-rule for those helicity PDFs. 
Polarized SIDIS is particularly important for the separate extraction of (anti-)quark helicity PDFs from data. 
This makes it a prominent observable to be measured at the upcoming Electron-Ion collider (EIC) at the Brookhaven National Laboratory~\cite{AbdulKhalek:2021gbh}.
The unique opportunities to study it at the EIC challenge the accuracy of available QCD theory predictions and provide motivation for their improvements, which will be addressed in this letter.

The reaction $l(k_l) + H(P) \rightarrow  l({k}'_l) + H'(P_H) + X$ defines the SIDIS process, 
where $k_l$, ${k}'_l$ ($P$, $P_H$) are momenta of incoming and outgoing leptons (hadrons), respectively, and 
the virtual photon momentum $q= k_l - {k}'_l$ squared, $Q^2=-q^2$, is large. 
The QCD improved parton model allows to express infrared safe observables in SIDIS through CFs, PDFs and FFs.
The hadron level cross section for unpolarized (spin averaged) SIDIS is given in terms of 
SFs $F_{1,2,3}$.
Exact results for the CFs of $F_{1,2}$ up to next-to-leading order (NLO) in perturbative QCD were obtained long ago~\cite{Altarelli:1979kv,Furmanski:1981cw} and the resummation of large threshold logarithms for SIDIS 
has been accomplished up to third order in QCD~\cite{Cacciari:2001cw,Anderle:2012rq,Anderle:2013lka,Abele:2021nyo,Abele:2022wuy}.
Recently, thanks to state-of-the-art theoretical developments in the computation of Feynman loop and phase-space integrals, the CFs have been computed to next-to-next-to-leading order (NNLO) accuracy. 
We have presented the first NNLO results (non-singlet parton channels and leading color approximation) in \cite{Goyal:2023zdi}. Subsequently, the complete results for the CFs of $F_{1,2}$ (all parton channels and full color dependence) 
have become available~\cite{Bonino:2024qbh,Goyal:2024emo} and both results agree with each other for all the channels.

Thus far, the description of polarized SIDIS in QCD has only been available at NLO  accuracy~\cite{deFlorian:1997zj}. 
In this letter we present, for the first time, the full NNLO QCD corrections. 
Polarized SIDIS is defined by the asymmetry
\begin{align*}
\frac{d^3 \Delta \sigma}{dx dy dz}  
 =
\frac{1}{2}\Bigg(
\frac{d^3 \sigma^{e^{-}_{\uparrow }H_{\uparrow}\rightarrow e^{-} H'X}}{dx dy dz}
-\frac{d^3 \sigma^{e^{-}_{\uparrow} H_{\downarrow}\rightarrow e^{-} H'X}}{dx dy dz}\Bigg)
\, ,
\end{align*}
where $e^{-}_{\uparrow }H_{\uparrow(\downarrow)}$ denote the (anti-)parallel spin-orientations of the colliding electron and hadron.
Here $x=\frac{Q^2}{2 P \cdot q}$ is the Bjorken variable, 
$y=\frac{P\cdot q}{P\cdot k_l}$ the inelasticity, 
and $z= \frac{P \cdot P_H}{P \cdot q}$ the scaling variable of the identified hadron.
The hadronic cross section above factorises into spin-dependent leptonic and hadronic tensors  $\Delta L_{\mu \nu}$
and $\Delta W_{\mu \nu}$,
\begin{equation}
\frac{d^3\Delta \sigma}{dxdydz} =\frac{2\pi y \alpha_e^{2}}{Q^4}\Delta L^{\mu\nu}(k_l,k'_l,q)\Delta W_{\mu\nu}(P,P_H,q)\, .
\end{equation}
Here $\Delta L^{\mu\nu}= -2 i \epsilon^{\mu \nu \sigma \lambda} q_\sigma s_{l,\lambda}$, 
with the spin vector $s_l$ of the incoming lepton, $\epsilon^{\mu\nu\sigma\lambda}$ is the Levi-Civita tensor (with $\epsilon^{0123} = -\epsilon_{0123} = -1$). The hadronic tensor $\Delta W_{\mu\nu}$ can be expressed in terms of spin-dependent SFs $g_1$ and $g_2$ as 
\begin{equation}
\Delta W_{\mu\nu} = g_1 (x,z,Q^2)T_{g_1,\mu \nu} + g_2 (x,z,Q^2) T_{g_2,\mu \nu}
\, ,
\end{equation}
with Lorentz tensors $T_{g_1,\mu\nu}$ = $\frac{i}{P.q} \epsilon_{\mu \nu \sigma \lambda} q^\sigma S^\lambda$ and 
$T_{g_2,\mu\nu}$= $\frac{i}{P.q} \epsilon_{\mu \nu \sigma \lambda} q^\sigma  (S^\lambda - \frac{S\cdot q}{P\cdot q }P^\lambda)$, and 
$S$ being the spin vector of the incoming hadron. 
For longitudinal polarization of the incoming hadron, $g_1$ is the dominant SF in the hadronic cross section,
\begin{equation}    
\frac{d^3\Delta \sigma}{dxdydz} = \frac{4\pi\alpha_e^2}{Q^2} \big(2-y\big)g_1 (x,z,Q^2)\, ,
\end{equation}
where $\alpha_e$ is the fine structure constant.  
With QCD factorization at scale $\mu_F$ the SF $g_1$ takes the form
\begin{align}
\label{eq:SFdef}
g_1 &= \sum_{a,b}\int_x^1 \frac{dx_1}{x_1} \Delta f_a(x_1,\mu_F^2) \int_z^1 \frac{dz_1}{z_1} D_b(z_1,\mu_F^2) 
\nonumber \\
    & \times \mathcal{G}_{1,ab}\left( \frac{x}{x_1}, \frac{z}{z_1}, \mu_F^2, Q^2\right )
\, ,
\end{align}
where $\Delta f_a$ = {$f_{a(\uparrow)/H(\uparrow)}
-f_{a(\downarrow)/H(\uparrow)} $ are the spin-dependent PDFs
and $D_b$ denote the spin-averaged FFs. 
Here the momentum fraction $x_1=\frac{p_a}{P}$ is carried by the initial parton `$a$' of incident hadron $H$ and $z_1=\frac{P_H}{p_b}$ by the hadron $H'$ with respect to the final state parton `$b$'.
The CFs $\mathcal{G}_{1,ab}$ are computable in perturbative QCD  
in powers of the strong coupling, $a_s(\mu_R^2) = {\alpha_s(\mu_R^2)}/(4\pi)$,
at the renormalization scale $\mu_R$,
\begin{equation}    
\label{eq:as-exp}
{\cal G}_{1,ab}(\mu_F^2) = \sum_{i=0}^\infty\, a_s^i(\mu_R^2)\, {\cal G}_{1,ab}^{(i)}(\mu_F^2,\mu_R^2)
\, ,
\end{equation}
where we have suppressed the scaling variables. 
$\mathcal{G}_{1,ab}$ is related to the parton level scattering cross sections  $d{\Delta \hat \sigma}_{1,ab}$ through projection with $\mathcal{P}^{\mu\nu}_{g_1}$, 
\begin{equation}
\label{eq:parton-crs}
d{\Delta \hat \sigma}_{1,ab} = \frac{\mathcal{P}_{g_1}^{\mu\nu}}{4\pi}\int \text{dPS}_{X+b}\, {\Sigma}|{\Delta M}_{ab}|^{2}_{\mu\nu}\, \delta\Big(\frac{z}{z_1}-\frac{p_a \cdot p_b}{p_a\cdot q}\Big) 
\end{equation}
where the projector in $D$ space-time dimensions reads, 
\begin{equation}
\mathcal{P}^{\mu\nu}_{g_1}= \frac{-i}{(D-2)(D-3)}
\epsilon^{\mu \nu \sigma \lambda} \,  \frac{q_\sigma p_{a,\lambda}}{p_a\cdot q} 
\, .
\end{equation}
$\Delta M_{ab} = M_{a(\uparrow)b}-M_{a(\downarrow)b}$  is the spin-dependent amplitude for the process $a(p_a,s_a)+\gamma^{*}(q)\rightarrow b(p_b) + X $, 
where the parton `$b$' fragments into hadron $H'$.  Here $s_a$ denotes the spin of the incoming parton $a$.  $\text{dPS}_{X+b}$ is the phase space for the final state particles consisting of $X$ and $b$.  
${\Sigma}$ denotes the summation over final state spin/polarization and their color quantum numbers 
in addition to the average over colors of incoming parton $a$.

At leading order (LO) in perturbation theory, the partonic cross sections in eq.~(\ref{eq:parton-crs}) receive a contribution from $\gamma^*+q (\bar{q}) \rightarrow q (\bar{q})$.  
At NLO, we consider one-loop corrections to the Born process $\gamma^*+q (\bar{q}) \rightarrow q (\bar{q})$, the real emission  $\gamma^* + q (\bar{q}) \rightarrow q (\bar{q}) + g$ 
and the gluon-initiated  $\gamma^* + g \rightarrow q + \bar{q}$ sub-processes.  
At NNLO, we include 
two-loop corrections to the Born process $\gamma^* + q(\bar{q}) \rightarrow q(\bar{q})$,
one-loop contributions to the single-gluon real emission $\gamma^* + q (\bar{q}) \rightarrow q (\bar{q}) + g$, 
and double real emissions $\gamma^* + q(\bar{q}) \rightarrow q(\bar{q}) + g + g$,
$\gamma^*+ g \rightarrow q + \overline q +g$ and
$\gamma^* + q(\bar{q}) \rightarrow q(\bar{q}) + q' + \bar{q}'$, 
where $q'$ can be of same or of different flavor as $q$.  
Note that in every sub-process, we need to include fragmentation contributions from each final state parton.

Beyond LO in perturbative QCD, we encounter both ultraviolet (UV) and 
infrared (IR) singularities.  The latter are due to the presence of soft and collinear partons.  
We regulate these singularities using dimensional regularization with $D=4+\varepsilon$ 
space-time dimensions.
The projection of spin-dependent partonic amplitudes squared $|\Delta M_{ab}|^2$ in eq.~(\ref{eq:parton-crs}) requires Dirac matrices $\gamma_5$ or the Levi-Civita tensor for 
polarized quarks or gluons, respectively, see, e.g. \cite{Zijlstra:1993sh}. 
Since $\gamma_5$ and the Levi-Civita tensor are intrinsically four-dimensional objects, their treatment in $D$ dimensions requires some prescription. 
Although, several schemes to define them in $D$ dimensions have been proposed, 
none of them is known to preserve the chiral Ward identity.  
A given prescription then requires an additional renormalization constant or an evanescent counter-term to preserve this identity.  
In this letter, we use Larin's prescription \cite{Larin:1993tq}  and replace $\slash\!\!\!p_a\gamma_5$ by ${\frac{i}{6}} ~\epsilon_{\mu \nu \sigma \lambda} p_a^\mu \gamma^\nu \gamma^\sigma \gamma^\lambda$.  The product of two 
Levi-Civita tensors 
is computed through  the determinant of Kronecker deltas defined in $D$ dimensions.     
The UV singularities are regulated through the renormalization of the strong coupling at the scale $\mu_R$.  
The IR singularities cancel among virtual and real emission processes, except those
from either incoming or tagged final state partons that are collinear to the rest of partons.   
Mass factorization guarantees that 
the partonic cross sections in eq.~(\ref{eq:parton-crs}) factorise into the spin-dependent Altarelli-Parisi (AP) kernels 
$\Delta \Gamma_{c\leftarrow a}$ of PDFs and $\tilde \Gamma_{b\leftarrow d}$ of FFs, 
appropriately convoluted with the finite CF (${\cal G}_{I,cd}$) 
at an arbitrary scale $\mu_F$ (suppressed here for brevity),  
\begin{eqnarray}
\label{eq:massfact}
{d\Delta \hat \sigma_{1,ab}(\varepsilon)}  = 
\Delta \Gamma_{c\leftarrow a}(\varepsilon) \otimes 
{\cal G}_{1,cd}(\varepsilon) \tilde \otimes 
\tilde \Gamma_{b\leftarrow d}(\varepsilon)
\, ,
\, 
\end{eqnarray}  
where summation over $c,d$ is implied and $\otimes$ 
($\tilde \otimes$) denotes a convolution over the scaling variable corresponding to PDFs (FFs), $x'=x/x_1$ ($z'=z/z_1$), cf. eq.~(\ref{eq:SFdef}).

The polarized space-like AP kernels ($\Delta \Gamma_{c\leftarrow a}$) are known at the order required~\cite{Mertig:1995ny,Vogelsang:1995vh,Vogelsang:1996im,Moch:2014sna,Blumlein:2021enk,Blumlein:2021ryt,Blumlein:2022gpp}.
Since the partonic cross sections in eq.(\ref{eq:massfact}) are derived in Larin's scheme, 
these spin-dependent AP kernels need to be taken in the same scheme, see \cite{Moch:2014sna}.
On the other hand, the spin-averaged time-like AP kernels ($\tilde \Gamma_{b\leftarrow d}$) are taken in the standard $\overline{\rm{MS}}$ scheme~\cite{Almasy:2011eq,Chen:2020uvt}.

The hadronic cross section (and the SF $g_1$) is independent of the prescription for $\gamma_5$.
Thus, QCD factorization allows to write $g_1$ in eq.~(\ref{eq:SFdef}) as 
\begin{align}
\label{eq:larinG}
g_1 = \sum_{a,b} \Delta f_{a,L}(\mu_F^2)\otimes  {\cal G}_{1,ab,L}(\mu_F^2) \tilde \otimes D_b(\mu_F^2)
\end{align}
where the subscript $L$ in $\Delta f_{a,L}$ and ${\cal G}_{1,ab,L}(\mu_F^2)$ 
denotes PDFs and CFs 
defined using Larin's scheme. 
It is straightforward to convert these quantities into $\overline{\rm{MS}}$ ones~\cite{Moch:2014sna}.  
The CFs in the ${\overline{\rm{MS}}}$ scheme are obtained by transforming 
$\Delta f_{a,L}$ to $\overline{\rm{MS}}$ PDFs through $\Delta f_{a}= Z_{ca}(\mu_F^2) \otimes f_{c,L}(\mu_F^2)$ and CFs to $\overline{\rm{MS}}$ CFs, ${\cal G}_{1,ab}= (Z^{-1}(\mu_F^2))_{ad}\otimes  {\cal G}_{1,db,L} (\mu_F^2)$.  
The finite renormalization constants $Z_{ab}$ are dependent on $x'$ and well known~\cite{Matiounine:1998re,Ravindran:2003gi,Moch:2014sna}.
We present the CFs in the $\overline{\rm{MS}}$ scheme in an ancillary file. 
The flavor-nonsinglet CFs of polarized SIDIS agree with those of the SF $F_3$, cf.~\cite{Goyal:2025xxx}.
The latter require a renormalization of the axial current and a kinematics independent finite renormalization from the Larin to the $\overline{\rm{MS}}$ scheme~\cite{Larin:1993tp,Ahmed:2015qpa}.
We find full agreement, which checks the scheme transformations applied.
Before we proceed to report the numerical impact of NNLO contributions to $g_1$, we briefly describe, how the cross sections $d \Delta \hat{\sigma}_{1,ab}$ in eq.~(\ref{eq:parton-crs}) are computed in Larin's scheme (denoted by $d \Delta \hat{\sigma}_{1,ab,L}$).

Beyond LO, the contributions to $d \Delta \hat{\sigma}_{1,ab,L}$ can be classified into three categories: pure virtual (VV),
pure real emissions (RR) and interference of real emission and virtual (RV). 
The VV part gets contributions from one-loop and two-loop virtual corrections to the Born process.  The latter can
be obtained using the quark form factor, see~\cite{Lee:2022nhh}.  
For the rest, we follow
the standard Feynman diagrammatic approach. We use \texttt{QGRAF}~\cite{Nogueira:1991ex} 
to generate Feynman diagrams and use a set of in-house routines written in 
\texttt{FORM} \cite{Kuipers:2012rf,Ruijl:2017dtg}, to convert the output of
\texttt{QGRAF} into a suitable format to apply Feynman rules and to perform Dirac algebra, Lorentz contractions and simplifications of color factors.  
The computations of phase-space integrals are
challenging compared to those required for inclusive cross sections because of the presence of an additional constraint 
$\frac{z}{z_{1}} = \frac{p_a \cdot p_b}{p_a\cdot q}$. 
The two-body phase-space over one-loop Feynman integrals that appear in RV and
three-body phase space integrals in RR are simplified with reverse unitarity~\cite{Anastasiou:2003gr,Anastasiou:2012kq}. 
This method allows us to apply loop-integration techniques, namely 
integration-by-parts identities (IBP) \cite{Chetyrkin:1981qh,Laporta:2001dd}, 
to reduce the phase-space integrals to a smaller number of the master integrals (MIs). 
The constraint $\frac{z}{z_{1}}= \frac{p_a \cdot p_b}{p_a \cdot q}$ is introduced through the delta function $\delta \left(z' - \frac{p_a \cdot p_b}{p_a \cdot q} \right)$, 
recall $z'=z/z_1$, which is replaced by a propagator-like term 
$-\frac{1}{\pi}\text{Im}(1/(z' - \frac{p_a \cdot p_b}{p_a \cdot q} + i \epsilon))$ 
with $p_b = p_a + q - k_1$ or $p_b = p_a + q - k_1 - k_2$ 
for two- and three-body final states respectively.
To perform the IBP reduction, we use the \texttt{Mathematica} package \texttt{LiteRed}~\cite{Lee:2013mka}.

After IBP reduction, we end up with 7 MIs for RV and 20 MIs for RR sub-processes.  
Due to the delta function constraint,
the results of the MIs depend on two scaling variables ($x', z'$). 
We have used two different approaches to compute these integrals.  
In the first approach, we choose a convenient Lorentz frame to parameterize the momenta so that  
the constraint on $z'$ takes the simple form and three-body phase-space integrals become three-dimensional parametric integrals, see~\cite{Matsuura:1988sm,Zijlstra:1992qd,Rijken:1996ns,Ravindran:2003um} 
for more details.  
We encounter two angular integrals and one parametric integral. 
Angular integrals reduce to
hypergeometric functions and the parametric integrals over these functions lead to multiple polylogarithms (MPLs) and Nielsen polylogarithms of weight up to three. 
In the second approach, we use the method of differential equations (DEs)~\cite{Kotikov:1990kg,Argeri:2007up,Remiddi:1997ny,Henn:2013pwa,Ablinger:2015tua} to solve the integrals.
We set up the system of differential equations of the MIs with respect to the variables 
$x',z'$ using \texttt{LiteRed}. 
Each set of DEs is controlled by a 20$\times$20 matrix. 
By an appropriate set of transformations on the set of MIs, we can
express these matrices in an upper or lower-triangular form
leading to the bottom-up approach of solving the DEs one by one. 
Alternatively, we use the elegant approach of an $\varepsilon$-factorized form~\cite{Henn:2013pwa} to reduce the DEs to canonical form with the help of the \texttt{Mathematica} package \texttt{Libra}~\cite{Lee:2020zfb}.
We use suitable boundary conditions to express the solution in terms of either classical polylogarithms or generalized harmonic polylogarithms (GPLs).
The boundary conditions for the MIs are computed in the threshold limit from parametric integrals.
We encounter four types of square-roots in the DE systems: $\big(\sqrt{x'}$, $\sqrt{z'}$, $\sqrt{(1+x')^2-4x'z'}$, $\sqrt{(1-z')^2+4x'z'}\big)$.
Thanks to suitable transformations on $x',z'$,
we can express all the polylogarithms or GPLs
with simple indices, ready for numerical evaluations.

The task to perform the mass factorization 
for the partonic cross sections in eq.~(\ref{eq:massfact}) 
to obtain finite CFs proceeds as follows. 
The AP kernels $\Delta \Gamma_{c\leftarrow a}$ and 
$\tilde \Gamma_{b\leftarrow d}$ in eq.~(\ref{eq:massfact}) are pure counter-terms, containing only poles in $\varepsilon$ in order to cancel the collinear singularities present in $d \Delta \hat{\sigma}_{1,ab,L}$. 
They contain standard `plus'-distributions ${\cal D}_j(w)=(\log^j(1-w)/(1-w))_+$ (see, e.g.~\cite{Goyal:2023zdi}) and delta functions $\delta(1-w)$,   
where $w=x',z'$, in addition to regular terms. 
The cancellation of the collinear singularities in $d \Delta \hat{\sigma}_{1,ab,L}$ against those from AP kernels requires to express the former ones in terms of the same distributions and regular functions. 
This is the most challenging task. 
In the partonic cross sections we encounter terms proportional to $(1-x')^{-1}$ and/or $(1-z')^{-1}$, 
which diverge in the respective threshold regions $x'\rightarrow 1$ and/or $z' \rightarrow 1$ respectively. 
These terms can originate either from MIs or their coefficients at the level of squared matrix elements.  
These singularities are regulated by $(1-x')^{a \varepsilon}$ and $(1-z')^{b\varepsilon}$ respectively
resulting from phase space and loop integrals.  
In addition we encounter spurious singularities when $x'=z'$ or $x'+z'=1$, which cancel among themselves at the end.  
In general, the resulting expressions contain multi-valued functions and we need to define them in different regions appropriately.  
We encounter different regions depending on whether $x'>z'$ with $x'+z'>1$ and/or $x'+z'<1$ or $x'<z'$ with $x'+z'>1$ and/or $x'+z'<1$.  
Using Feynman's $i \epsilon$ prescription, we can analytically continue these functions smoothly from one region to other.

E.g., in the RV sub-processes, we encounter a hypergeometric function which after Pfaff transformation gives 
\begin{align}\label{eq:hyper}
&{}_{2}F_{1}\left(1,\frac{\varepsilon}{2};1+\frac{\varepsilon}{2};\frac{x'(1-x')}{z'(1-z')}\right) 
   = \bigg[\frac{(z'-x')(1-z'-x')}{z'(1-z')}\bigg]^{-\frac{\varepsilon}{2}} \nonumber \\
   & \hspace{1.2cm}\times {}_{2}F_{1}\left(\frac{\varepsilon}{2},\frac{\varepsilon}{2};1+\frac{\varepsilon}{2};\frac{x'(1-x')}{(x'-z')(1-z'-x')}\right)
\end{align}
Using   Feynman's $i\epsilon$ prescription for scaling variables, i.e. $x'\equiv x'-i\epsilon$ and $z'\equiv z'-i\epsilon$ and the identities involving theta functions, eq.~(\ref{eq:hyper}) reduces to
\begin{align}
&A~\bigg(\theta_{1a}\Big(\theta_{2a}\ F(\varepsilon,B+i\epsilon)+\theta_{2b}\ (-1+i\epsilon)^{-\frac{\varepsilon}{2}}F(\varepsilon,B+i\epsilon)  \Big)\nonumber\\
&+\hspace{0.2cm}\theta_{1b}\Big(\theta_{2b}\ F(\varepsilon,B -i\epsilon)+\theta_{2a}\ (-1-i\epsilon)^{-\frac{\varepsilon}{2}}
F(\varepsilon,B-i\epsilon)  \Big)\bigg)\nonumber
\end{align}
where, $\theta_{1a} = \theta(z'-x')$, $\theta_{1b} = \theta(x'-z')$, $\theta_{2a} = \theta(1-z'-x')$, $\theta_{2b} = \theta(z'+x'-1)$, $A$=$\Big|\frac{(z'-x')(1-z'-x')}{z'(1-z')}\Big|^{-\frac{\varepsilon}{2}}$, $ F(\varepsilon,y)= {}_{2}F_{1}(\frac{\varepsilon}{2}, \frac{\varepsilon}{2};1+\frac{\varepsilon}{2};y)$ and $B = \frac{x'(1-x')}{(x'-z')(1-z'-x')}$.
$F(\varepsilon,B\pm i\epsilon)$ can be analytically continued to the appropriate region and expanded in a power series in $\varepsilon$, see \cite{Duplancic:2000sk,Gehrmann:2002zr,Gehrmann:2022cih}. Finally, collinear singularities in $w=x',z'$ are exposed through 
\begin{align}
(1-w)^{-1 + n \varepsilon} = \frac{1}{n \varepsilon} \delta(1-w) + \sum_{k=0}^\infty \frac{(n \varepsilon)^k}{k!}  \mathcal{D}_{k}(w) \, .
\end{align}

The resulting partonic cross sections $d \Delta \hat{\sigma}_{1,ab,L}$ contain
double and single poles in $\varepsilon$ at NLO. 
The former ones cancels between VV and RR terms and the latter against AP kernels in the mass factorization eq.~(\ref{eq:massfact}). 
At NNLO the leading $1/\varepsilon^4$ and $1/\varepsilon^3$ poles cancel among  the VV, RV and RR contributions. 
The remaining double and single poles in $\varepsilon$ cancel against the AP kernels using eq.~(\ref{eq:massfact}). 
The final $\overline{\rm{MS}}$ scheme CFs thus obtained 
(after transformation from the Larin scheme) can be written as, 
\begin{eqnarray}
\label{eq:CFsvnsv}
{\cal G}_{1,ab} &=&  \sum_{r}
\Delta C_{ab}^r  h_r (x',z')  
+ \sum_{\beta} \Big({\Delta C }_{ab,x'}^\beta(x') Z_\beta(z')\nonumber \\
&&+ {\Delta C}_{ab,z'}^\beta(z') X_\beta(x')\Big)
+ \Delta R_{ab}(x',z')
\, .
\end{eqnarray}
The soft plus virtual (SV) terms $h_r$ contain the double distributions 
$h_{\delta_{x'} \delta_{z'}}=\delta(1-x') \delta(1-z')$, 
$h_{\delta_{x'} j}= \delta(1-x') {\cal D}_j(z')$, 
$h_{j, \delta_{z'}} = {\cal D}_j(x') \delta(1-z')$,   
$h_{j k}=  {\cal D}_j(x') {\cal D}_k(z')$. 
Terms with single distributions, namely 
$Z_{\delta_{z'}}(z') =\delta(1-z')$,  
$Z_{j}(z') ={\cal D}_j(z')$, 
$X_{\delta_{x'}}(x')=\delta(1-x')$, 
$X_j(x')={\cal D}_j(x') $ 
are called partial-SV (pSV) terms and regular terms are denoted by $\Delta R_{ab}$. Our NLO results are in complete agreement with \cite{deFlorian:1997zj}.
$\Delta C_{ab}^r $ are in complete agreement with those of unpolarized SFs $F_{1,2}$ \cite{Abele:2021nyo,Goyal:2023zdi,Bonino:2024qbh,Goyal:2024emo}, see also \cite{Ravindran:2006bu,Ahmed:2014uya} and if we expand ${\Delta C }_{ab,x'}^\beta(x')$ and  ${\Delta C}_{ab,z'}^\beta(z')$ around $x', z' \rightarrow 1$, the results are in complete agreement with the corresponding terms in the unpolarized case up to order $(1-x')^{0}$ and $(1-z')^{0}$ respectively, \cite{Bonino:2024qbh,Goyal:2024emo,AH:2020qoa}. 
The remaining contributions in eq.~(\ref{eq:CFsvnsv}), i.e., single distributions $X_\beta$, $Z_\beta$
and regular terms $\Delta R_{ab}$ are new. 
These results are too lengthy to be presented here and, instead included in an ancillary file.

In the following, we illustrate the numerical impact of  our results for ${\cal G}_{1,ab}$ for various centre-of-mass energies $\sqrt{s}$ for the range of $x$ and $z$ values.  
The convolution of the CFs with PDFs and FFs provides $g_1 = \sum_{i=0} a_s^i g_1^{(i)}$, such that 
at LO $g_1^{(0)}=\sum_{q} e_{q}^{2} H_{qq}$ ($e_q$ being the electric charge of quark $q$).
\begin{align}
\label{eq:nlo-g1}
g_1^{(1)} & = \sum_{q} e_{q}^{2}\bigg(  H_{qq}\hat \otimes G_{1,qq}^{(1)} + H_{qg} \hat \otimes G_{1,qg}^{(1)} + H_{gq} \hat \otimes G_{1,gq}^{(1)} \bigg) 
\, ,
\end{align}
\begin{widetext}
\begin{align}
\label{eq:nnlo-g1}
g_1^{(2)} &= \sum_{q} e_{q}^{2}\bigg(  H_{qq}\hat \otimes G_{1,qq,\text{NS}}^{(2)} + H_{q\bar{q}}\hat \otimes  G_{1,q\bar{q}}^{(2)} + H_{qg} \hat \otimes G_{1,qg}^{(2)} + H_{gq} \hat \otimes G_{1,gq}^{(2)} \bigg) +
\Big(\sum_{q_i} e_{q_i}^{2}\Big)\bigg( H_{qq}\hat \otimes G_{1,qq,\text{PS}}^{(2)}  + H_{gg} \hat \otimes G_{1,gg}^{(2)} \bigg) \nonumber\\
&+ \sum_{q} \sum_{q'\neq q}\bigg( e_{q}^{2}~H^{+}_{qq'} \hat \otimes  G_{1,qq',[1]}^{(2)} + e_{q'}^{2}~H^{+}_{qq'}\hat \otimes  G_{1,qq',[2]}^{(2)} + e_{q}e_{q'} H^{-}_{qq'} \hat \otimes G_{1,qq',[3]}^{(2)}\bigg)
\, .
\end{align}
\end{widetext} 
The ${G}^{(i)}_{1,ab}$ are related to ${\cal G}^{(i)}_{1,ab}$ defined in eq.~(\ref{eq:as-exp}), see comments in ancillary file, and  $\hat\otimes$ denotes their convolution with $H_{ab}$ in both variables $x$ and $z$.
\begin{align}
H_{qq} &= \Delta f_q(x) D_q(z) + \Delta f_{\bar{q}} (x)D_{\bar{q}}(z)\, , \nonumber\\ 
H_{q\bar{q}} &= \Delta f_q (x) D_{\bar{q}} (z)+ \Delta f_{\bar{q}}(x) D_q(z)\, , \nonumber\\
H_{qg} &= \Delta f_q (x) D_g (z)+\Delta f_{\bar{q}}(x) D_g (z)\, , \nonumber\\
H_{gg} &= \Delta f_g (x)D_g(z)\, , \nonumber\\
H^{\pm}_{qq'} &= \Delta f_q (x)D_{q'}(z) \pm \Delta f_q (x)D_{\bar{q}'}(z) \nonumber\\ 
& \qquad \pm  \Delta f_{\bar{q}} (x)D_{q'}(z)+  \Delta f_{\bar{q}} (x) D_{\bar{q}'}(z)\, , \nonumber\\
H_{gq} &= \Delta f_g (x) D_q (z) +  \Delta f_g D_{\bar{q}}(z)\, .  
\end{align}
We evaluate eqs.~(\ref{eq:nlo-g1}) and (\ref{eq:nnlo-g1}) numerically for the $g_1^{\pi^{+}}$ SF (with an identified $\pi^{+}$ in the final state) as a function of $x$ and $Q^2$ after integrating $z$ over a range from 0.2 to 0.85.  We use polarized PDFs from \texttt{MAPPDF10NLO} at LO and NLO and \texttt{MAPPDF10NNLO} at NNLO throughout~\cite{Bertone:2024taw}.
The strong coupling constant  $\alpha_s$ is taken from the PDF sets  at the respective perturbative order and use $n_F=3$ active flavors. 
\begin{figure}[!ht]
\includegraphics[width=0.465\textwidth]{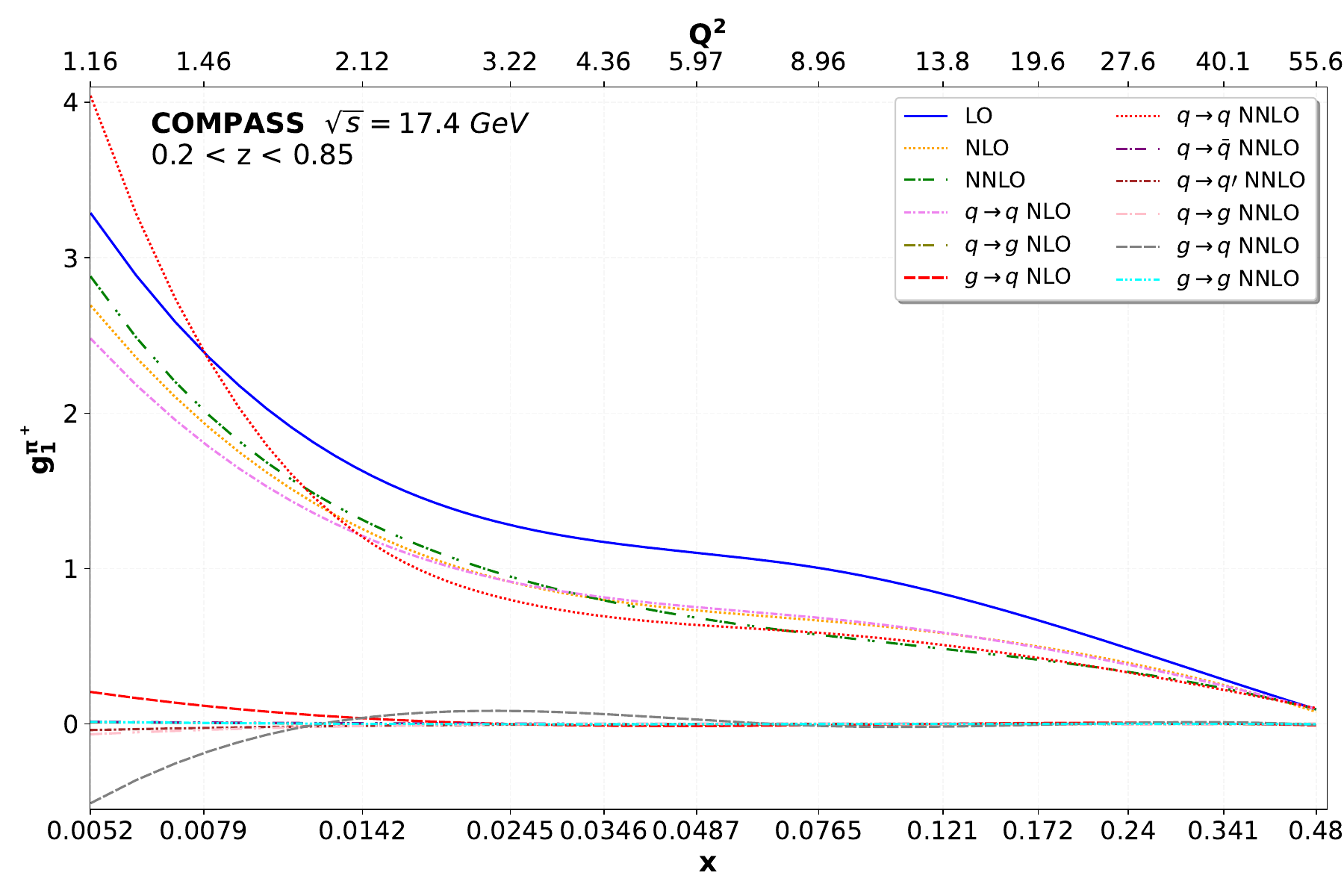}
\caption{Contributions from all partonic channels to SF $g_1^{\pi^{+}}$ as a function of $x$ for the COMPASS energy $\sqrt{s}=17.4$ GeV.}
\label{fig:1}
\end{figure}
In Fig.~\ref{fig:1}, we show contributions to $g_1^{\pi^{+}}$  from all the partonic channels at various perturbative orders at $\sqrt{s}=17.4$ GeV after setting the scales 
$\mu_R^2$ = $\mu_F^2$ = $Q^2$, 
and using FFs from $\texttt{NNFF10}$~\cite{Bertone:2017tyb} at the respective perturbative orders. 
We observe at every order, that the partonic channel where incoming as well as fragmenting states are (anti-)quarks dominate over the rest of the partonic channels.  

In the Fig.~\ref{fig:2}, we show how $g_1^{\pi^{+}}$ changes with respect to renormalization and factorization scales at various perturbative orders as a function of $x$ at $\sqrt{s}=45$ GeV.  
In each plot, the range of $x$ is chosen by constraining $y=Q^2/(x s)$ to 
the range $0.5 \leq y \leq 0.9$.  
The bands at every order are the result of a 7-point scale variation around the central scale ($\mu_R^{2}=\mu_F^{2}=Q^2$) 
with $k_1\mu_R^{2}, k_2\mu_F^{2}$ where ($k_1, k_2$) $\in$ [1/2, 2] with a constraint 1/2 $\le$ $k_1/k_2 \le$ 2.  
We have used FFs  from \texttt{MAPFF10NLO} at LO and NLO, \texttt{MAPFF10NNLO} at NNLO~\cite{AbdulKhalek:2022laj}. 
It demonstrates clearly that the inclusion of NNLO corrections reduces the scale dependence when compared to the previous orders.
We find at small $Q^2$ the sensitivity to these scales is larger compared to that at larger values of $Q^2$.  The reason for this is that at small $Q^2$ the strong coupling is large and the effect of scale variations is, therefore, amplified.  
We also observe that the bands from NLO and NNLO predictions are well separated at lower values of $Q^2$.  The reason for this is that contributions from partonic channels  at order $\alpha_s^2$ (NNLO) are as large or even bigger than those at $\alpha_s$ (NLO).

In Fig.~\ref{fig:3}, we plot the ratio of the polarized SF $g_1^{\pi^{+}}$ to the unpolarized SF $F_1^{\pi^{+}}$ 
as a function of $x$ for the COMPASS energy $\sqrt{s}=17.4$ GeV, 
using $\texttt{NNPDF3.1}$~\cite{NNPDF:2017mvq} for unpolarized PDFs and $\texttt{NNFF10}$ for   FFs  at the respective perturbative orders.
The plot includes the 7-point scale variation at LO, NLO and NNLO and, for comparison, the experimental data taken from \cite{COMPASS:2010hwr}.
Given the accuracy of those data, Fig.~\ref{fig:3} clearly indicates the need to reduce residual theory uncertainties in SIDIS predictions at NNLO accuracy.

\begin{figure}[!ht]
\includegraphics[width=0.49\textwidth]{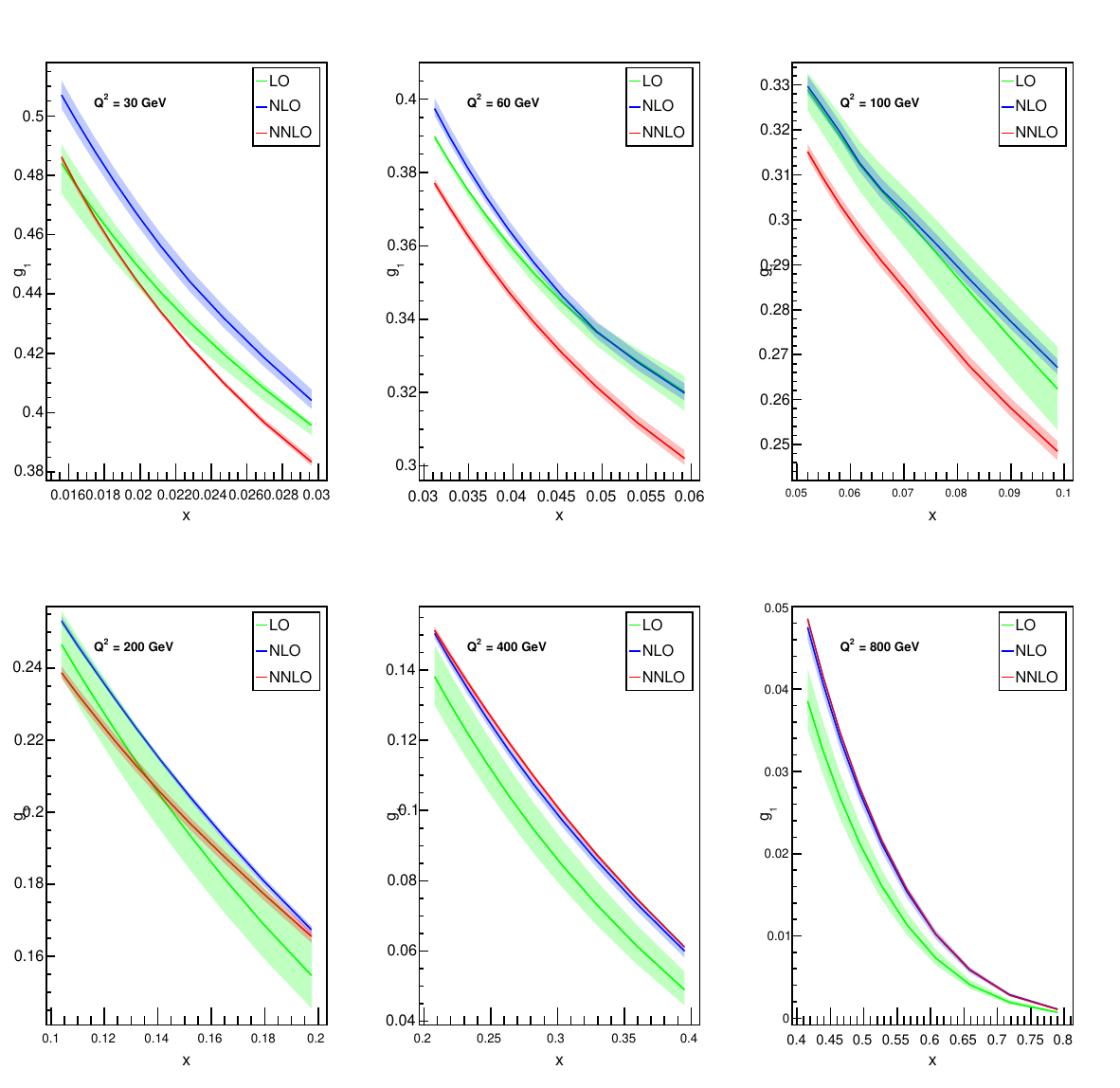}
\caption{Dependence of $g_1^{\pi^{+}}$ on renormalization and factorization scales 
in 7-point variation of $\mu_R^{2}$ and $\mu_F^{2}$, 
as a function of $x$ at various values of $Q^2$. }
\label{fig:2}
\end{figure}
\begin{figure}[!ht]
\includegraphics[width=0.51\textwidth]{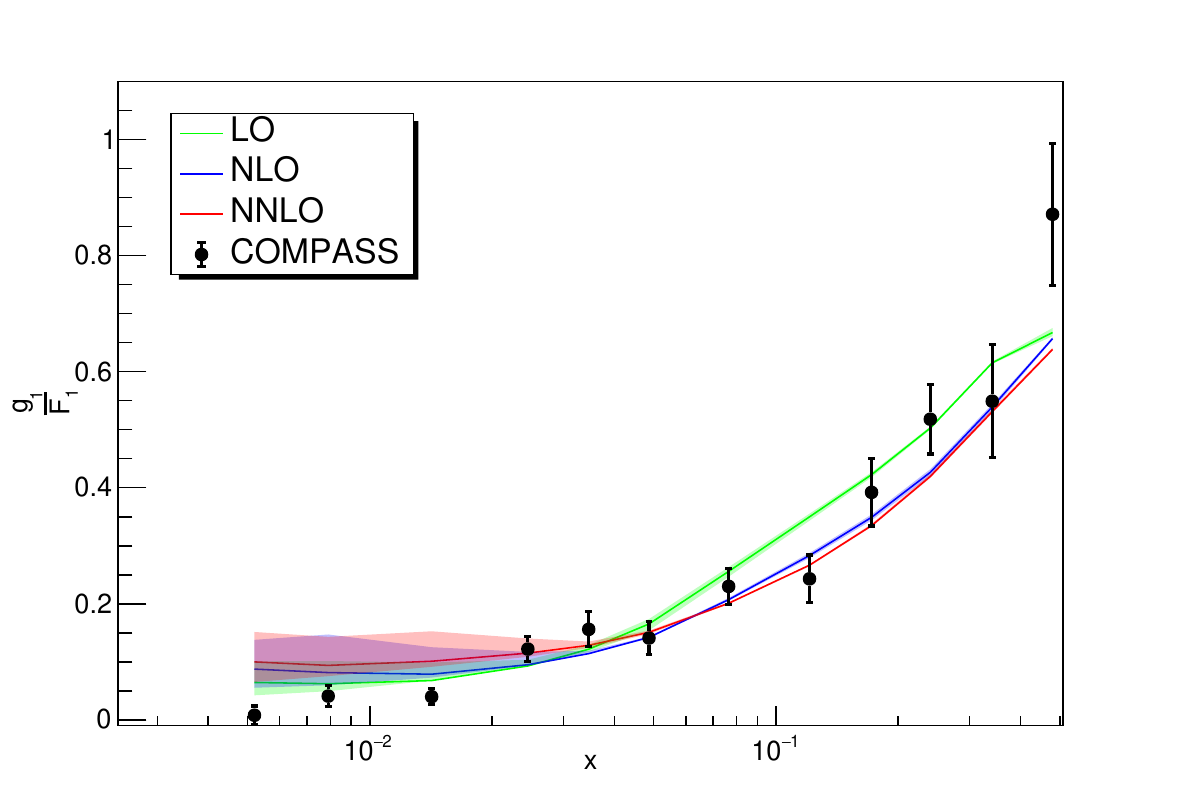}
\caption{ The ratio of SFs 
$g_1^{\pi^{+}}/F_1^{\pi^{+}}$ as a function of $x$ for the COMPASS energy $\sqrt{s}=17.4$ GeV.  
The bands are due to 7-point scale variation.}
\label{fig:3}
\end{figure}
In this letter, we report the CFs for the polarized SIDIS process at NNLO in QCD. 
These results close a prominent gap in the available literature. They will facilitate high precision theory predictions and will contribute to the studies of polarized PDFs and of the proton spin structure at the future EIC.
A \texttt{Mathematica} notebook with all results for the CFs ${\cal G}_{1,ab}^{(i)}$ is available from the preprint server {\tt https://arXiv.org}.

\begin{acknowledgments}
\emph{Acknowledgements:}  
We thank W.~Vogelsang and S.~Weinzierl for discussions. 
We thank the authors of \cite{Bonino:2024wgg,Bonino:2025bqa} for correspondence regarding the finite renormalization constant $Z_{ab}$.
The phenomenological results presented in the original paper are unaffected, as numerical impact of the modifications is not numerically significant.
This work has been supported through a joint Indo-German research grant by
the Department of Science and Technology (DST/INT/DFG/P-03/2021/dtd.12.11.21). 
S.M. acknowledges the ERC Advanced Grant 101095857 {\it Conformal-EIC}. The work of R.L. was supported via RSF grant No. 20-12-00205.
\end{acknowledgments}

\bibliographystyle{apsrev}
\bibliography{main}

\begin{thebibliography}{66}
\expandafter\ifx\csname natexlab\endcsname\relax\def\natexlab#1{#1}\fi
\expandafter\ifx\csname bibnamefont\endcsname\relax
  \def\bibnamefont#1{#1}\fi
\expandafter\ifx\csname bibfnamefont\endcsname\relax
  \def\bibfnamefont#1{#1}\fi
\expandafter\ifx\csname citenamefont\endcsname\relax
  \def\citenamefont#1{#1}\fi
\expandafter\ifx\csname url\endcsname\relax
  \def\url#1{\texttt{#1}}\fi
\expandafter\ifx\csname urlprefix\endcsname\relax\def\urlprefix{URL }\fi
\providecommand{\bibinfo}[2]{#2}
\providecommand{\eprint}[2][]{\url{#2}}

\bibitem[{\citenamefont{Bl{\"u}mlein}(2013)}]{Blumlein:2012bf}
\bibinfo{author}{\bibfnamefont{J.}~\bibnamefont{Bl{\"u}mlein}},
  \bibinfo{journal}{Prog. Part. Nucl. Phys.} \textbf{\bibinfo{volume}{69}},
  \bibinfo{pages}{28} (\bibinfo{year}{2013}), \eprint{1208.6087}.

\bibitem[{\citenamefont{Workman and Others}(2022)}]{Workman:2022ynf}
\bibinfo{author}{\bibfnamefont{R.~L.} \bibnamefont{Workman}} \bibnamefont{and}
  \bibinfo{author}{\bibnamefont{Others}} (\bibinfo{collaboration}{Particle Data
  Group}), \bibinfo{journal}{PTEP} \textbf{\bibinfo{volume}{2022}},
  \bibinfo{pages}{083C01} (\bibinfo{year}{2022}).

\bibitem[{\citenamefont{Metz and Vossen}(2016)}]{Metz:2016swz}
\bibinfo{author}{\bibfnamefont{A.}~\bibnamefont{Metz}} \bibnamefont{and}
  \bibinfo{author}{\bibfnamefont{A.}~\bibnamefont{Vossen}},
  \bibinfo{journal}{Prog. Part. Nucl. Phys.} \textbf{\bibinfo{volume}{91}},
  \bibinfo{pages}{136} (\bibinfo{year}{2016}), \eprint{1607.02521}.

\bibitem[{\citenamefont{Aidala et~al.}(2013)\citenamefont{Aidala, Bass, Hasch,
  and Mallot}}]{Aidala:2012mv}
\bibinfo{author}{\bibfnamefont{C.~A.} \bibnamefont{Aidala}},
  \bibinfo{author}{\bibfnamefont{S.~D.} \bibnamefont{Bass}},
  \bibinfo{author}{\bibfnamefont{D.}~\bibnamefont{Hasch}}, \bibnamefont{and}
  \bibinfo{author}{\bibfnamefont{G.~K.} \bibnamefont{Mallot}},
  \bibinfo{journal}{Rev. Mod. Phys.} \textbf{\bibinfo{volume}{85}},
  \bibinfo{pages}{655} (\bibinfo{year}{2013}), \eprint{1209.2803}.

\bibitem[{\citenamefont{de~Florian et~al.}(2009)\citenamefont{de~Florian,
  Sassot, Stratmann, and Vogelsang}}]{deFlorian:2009vb}
\bibinfo{author}{\bibfnamefont{D.}~\bibnamefont{de~Florian}},
  \bibinfo{author}{\bibfnamefont{R.}~\bibnamefont{Sassot}},
  \bibinfo{author}{\bibfnamefont{M.}~\bibnamefont{Stratmann}},
  \bibnamefont{and}
  \bibinfo{author}{\bibfnamefont{W.}~\bibnamefont{Vogelsang}},
  \bibinfo{journal}{Phys. Rev. D} \textbf{\bibinfo{volume}{80}},
  \bibinfo{pages}{034030} (\bibinfo{year}{2009}), \eprint{0904.3821}.

\bibitem[{\citenamefont{de~Florian et~al.}(2014)\citenamefont{de~Florian,
  Sassot, Stratmann, and Vogelsang}}]{deFlorian:2014yva}
\bibinfo{author}{\bibfnamefont{D.}~\bibnamefont{de~Florian}},
  \bibinfo{author}{\bibfnamefont{R.}~\bibnamefont{Sassot}},
  \bibinfo{author}{\bibfnamefont{M.}~\bibnamefont{Stratmann}},
  \bibnamefont{and}
  \bibinfo{author}{\bibfnamefont{W.}~\bibnamefont{Vogelsang}},
  \bibinfo{journal}{Phys. Rev. Lett.} \textbf{\bibinfo{volume}{113}},
  \bibinfo{pages}{012001} (\bibinfo{year}{2014}), \eprint{1404.4293}.

\bibitem[{\citenamefont{Nocera et~al.}(2014)\citenamefont{Nocera, Ball, Forte,
  Ridolfi, and Rojo}}]{Nocera:2014gqa}
\bibinfo{author}{\bibfnamefont{E.~R.} \bibnamefont{Nocera}},
  \bibinfo{author}{\bibfnamefont{R.~D.} \bibnamefont{Ball}},
  \bibinfo{author}{\bibfnamefont{S.}~\bibnamefont{Forte}},
  \bibinfo{author}{\bibfnamefont{G.}~\bibnamefont{Ridolfi}}, \bibnamefont{and}
  \bibinfo{author}{\bibfnamefont{J.}~\bibnamefont{Rojo}}
  (\bibinfo{collaboration}{NNPDF}), \bibinfo{journal}{Nucl. Phys. B}
  \textbf{\bibinfo{volume}{887}}, \bibinfo{pages}{276} (\bibinfo{year}{2014}),
  \eprint{1406.5539}.

\bibitem[{\citenamefont{Abdul~Khalek
  et~al.}(2022{\natexlab{a}})}]{AbdulKhalek:2021gbh}
\bibinfo{author}{\bibfnamefont{R.}~\bibnamefont{Abdul~Khalek}}
  \bibnamefont{et~al.}, \bibinfo{journal}{Nucl. Phys. A}
  \textbf{\bibinfo{volume}{1026}}, \bibinfo{pages}{122447}
  (\bibinfo{year}{2022}{\natexlab{a}}), \eprint{2103.05419}.

\bibitem[{\citenamefont{Altarelli et~al.}(1979)\citenamefont{Altarelli, Ellis,
  Martinelli, and Pi}}]{Altarelli:1979kv}
\bibinfo{author}{\bibfnamefont{G.}~\bibnamefont{Altarelli}},
  \bibinfo{author}{\bibfnamefont{R.~K.} \bibnamefont{Ellis}},
  \bibinfo{author}{\bibfnamefont{G.}~\bibnamefont{Martinelli}},
  \bibnamefont{and} \bibinfo{author}{\bibfnamefont{S.-Y.} \bibnamefont{Pi}},
  \bibinfo{journal}{Nucl. Phys. B} \textbf{\bibinfo{volume}{160}},
  \bibinfo{pages}{301} (\bibinfo{year}{1979}).

\bibitem[{\citenamefont{Furmanski and Petronzio}(1982)}]{Furmanski:1981cw}
\bibinfo{author}{\bibfnamefont{W.}~\bibnamefont{Furmanski}} \bibnamefont{and}
  \bibinfo{author}{\bibfnamefont{R.}~\bibnamefont{Petronzio}},
  \bibinfo{journal}{Z. Phys. C} \textbf{\bibinfo{volume}{11}},
  \bibinfo{pages}{293} (\bibinfo{year}{1982}).

\bibitem[{\citenamefont{Cacciari and Catani}(2001)}]{Cacciari:2001cw}
\bibinfo{author}{\bibfnamefont{M.}~\bibnamefont{Cacciari}} \bibnamefont{and}
  \bibinfo{author}{\bibfnamefont{S.}~\bibnamefont{Catani}},
  \bibinfo{journal}{Nucl. Phys. B} \textbf{\bibinfo{volume}{617}},
  \bibinfo{pages}{253} (\bibinfo{year}{2001}), \eprint{hep-ph/0107138}.

\bibitem[{\citenamefont{Anderle
  et~al.}(2013{\natexlab{a}})\citenamefont{Anderle, Ringer, and
  Vogelsang}}]{Anderle:2012rq}
\bibinfo{author}{\bibfnamefont{D.~P.} \bibnamefont{Anderle}},
  \bibinfo{author}{\bibfnamefont{F.}~\bibnamefont{Ringer}}, \bibnamefont{and}
  \bibinfo{author}{\bibfnamefont{W.}~\bibnamefont{Vogelsang}},
  \bibinfo{journal}{Phys. Rev. D} \textbf{\bibinfo{volume}{87}},
  \bibinfo{pages}{034014} (\bibinfo{year}{2013}{\natexlab{a}}),
  \eprint{1212.2099}.

\bibitem[{\citenamefont{Anderle
  et~al.}(2013{\natexlab{b}})\citenamefont{Anderle, Ringer, and
  Vogelsang}}]{Anderle:2013lka}
\bibinfo{author}{\bibfnamefont{D.~P.} \bibnamefont{Anderle}},
  \bibinfo{author}{\bibfnamefont{F.}~\bibnamefont{Ringer}}, \bibnamefont{and}
  \bibinfo{author}{\bibfnamefont{W.}~\bibnamefont{Vogelsang}},
  \bibinfo{journal}{Phys. Rev. D} \textbf{\bibinfo{volume}{87}},
  \bibinfo{pages}{094021} (\bibinfo{year}{2013}{\natexlab{b}}),
  \eprint{1304.1373}.

\bibitem[{\citenamefont{Abele et~al.}(2021)\citenamefont{Abele, de~Florian, and
  Vogelsang}}]{Abele:2021nyo}
\bibinfo{author}{\bibfnamefont{M.}~\bibnamefont{Abele}},
  \bibinfo{author}{\bibfnamefont{D.}~\bibnamefont{de~Florian}},
  \bibnamefont{and}
  \bibinfo{author}{\bibfnamefont{W.}~\bibnamefont{Vogelsang}},
  \bibinfo{journal}{Phys. Rev. D} \textbf{\bibinfo{volume}{104}},
  \bibinfo{pages}{094046} (\bibinfo{year}{2021}), \eprint{2109.00847}.

\bibitem[{\citenamefont{Abele et~al.}(2022)\citenamefont{Abele, de~Florian, and
  Vogelsang}}]{Abele:2022wuy}
\bibinfo{author}{\bibfnamefont{M.}~\bibnamefont{Abele}},
  \bibinfo{author}{\bibfnamefont{D.}~\bibnamefont{de~Florian}},
  \bibnamefont{and}
  \bibinfo{author}{\bibfnamefont{W.}~\bibnamefont{Vogelsang}},
  \bibinfo{journal}{Phys. Rev. D} \textbf{\bibinfo{volume}{106}},
  \bibinfo{pages}{014015} (\bibinfo{year}{2022}), \eprint{2203.07928}.

\bibitem[{\citenamefont{Goyal et~al.}(2024{\natexlab{a}})\citenamefont{Goyal,
  Moch, Pathak, Rana, and Ravindran}}]{Goyal:2023zdi}
\bibinfo{author}{\bibfnamefont{S.}~\bibnamefont{Goyal}},
  \bibinfo{author}{\bibfnamefont{S.-O.} \bibnamefont{Moch}},
  \bibinfo{author}{\bibfnamefont{V.}~\bibnamefont{Pathak}},
  \bibinfo{author}{\bibfnamefont{N.}~\bibnamefont{Rana}}, \bibnamefont{and}
  \bibinfo{author}{\bibfnamefont{V.}~\bibnamefont{Ravindran}},
  \bibinfo{journal}{Phys. Rev. Lett.} \textbf{\bibinfo{volume}{132}},
  \bibinfo{pages}{251902} (\bibinfo{year}{2024}{\natexlab{a}}),
  \eprint{2312.17711}.

\bibitem[{\citenamefont{Bonino et~al.}(2024{\natexlab{a}})\citenamefont{Bonino,
  Gehrmann, and Stagnitto}}]{Bonino:2024qbh}
\bibinfo{author}{\bibfnamefont{L.}~\bibnamefont{Bonino}},
  \bibinfo{author}{\bibfnamefont{T.}~\bibnamefont{Gehrmann}}, \bibnamefont{and}
  \bibinfo{author}{\bibfnamefont{G.}~\bibnamefont{Stagnitto}},
  \bibinfo{journal}{Phys. Rev. Lett.} \textbf{\bibinfo{volume}{132}},
  \bibinfo{pages}{251901} (\bibinfo{year}{2024}{\natexlab{a}}),
  \eprint{2401.16281}.

\bibitem[{\citenamefont{Goyal et~al.}(2024{\natexlab{b}})\citenamefont{Goyal,
  Lee, Moch, Pathak, Rana, and Ravindran}}]{Goyal:2024emo}
\bibinfo{author}{\bibfnamefont{S.}~\bibnamefont{Goyal}},
  \bibinfo{author}{\bibfnamefont{R.~N.} \bibnamefont{Lee}},
  \bibinfo{author}{\bibfnamefont{S.-O.} \bibnamefont{Moch}},
  \bibinfo{author}{\bibfnamefont{V.}~\bibnamefont{Pathak}},
  \bibinfo{author}{\bibfnamefont{N.}~\bibnamefont{Rana}}, \bibnamefont{and}
  \bibinfo{author}{\bibfnamefont{V.}~\bibnamefont{Ravindran}}
  (\bibinfo{year}{2024}{\natexlab{b}}), \eprint{2412.19309}.

\bibitem[{\citenamefont{de~Florian et~al.}(1998)\citenamefont{de~Florian,
  Stratmann, and Vogelsang}}]{deFlorian:1997zj}
\bibinfo{author}{\bibfnamefont{D.}~\bibnamefont{de~Florian}},
  \bibinfo{author}{\bibfnamefont{M.}~\bibnamefont{Stratmann}},
  \bibnamefont{and}
  \bibinfo{author}{\bibfnamefont{W.}~\bibnamefont{Vogelsang}},
  \bibinfo{journal}{Phys. Rev. D} \textbf{\bibinfo{volume}{57}},
  \bibinfo{pages}{5811} (\bibinfo{year}{1998}), \eprint{hep-ph/9711387}.

\bibitem[{\citenamefont{Zijlstra and van Neerven}(1994)}]{Zijlstra:1993sh}
\bibinfo{author}{\bibfnamefont{E.~B.} \bibnamefont{Zijlstra}} \bibnamefont{and}
  \bibinfo{author}{\bibfnamefont{W.~L.} \bibnamefont{van Neerven}},
  \bibinfo{journal}{Nucl. Phys. B} \textbf{\bibinfo{volume}{417}},
  \bibinfo{pages}{61} (\bibinfo{year}{1994}), \bibinfo{note}{[Erratum:
  Nucl.Phys.B 426, 245 (1994), Erratum: Nucl.Phys.B 773, 105--106 (2007),
  Erratum: Nucl.Phys.B 501, 599--599 (1997)]}.

\bibitem[{\citenamefont{Larin}(1993)}]{Larin:1993tq}
\bibinfo{author}{\bibfnamefont{S.~A.} \bibnamefont{Larin}},
  \bibinfo{journal}{Phys. Lett. B} \textbf{\bibinfo{volume}{303}},
  \bibinfo{pages}{113} (\bibinfo{year}{1993}), \eprint{hep-ph/9302240}.

\bibitem[{\citenamefont{Mertig and van Neerven}(1996)}]{Mertig:1995ny}
\bibinfo{author}{\bibfnamefont{R.}~\bibnamefont{Mertig}} \bibnamefont{and}
  \bibinfo{author}{\bibfnamefont{W.~L.} \bibnamefont{van Neerven}},
  \bibinfo{journal}{Z. Phys. C} \textbf{\bibinfo{volume}{70}},
  \bibinfo{pages}{637} (\bibinfo{year}{1996}), \eprint{hep-ph/9506451}.

\bibitem[{\citenamefont{Vogelsang}(1996{\natexlab{a}})}]{Vogelsang:1995vh}
\bibinfo{author}{\bibfnamefont{W.}~\bibnamefont{Vogelsang}},
  \bibinfo{journal}{Phys. Rev. D} \textbf{\bibinfo{volume}{54}},
  \bibinfo{pages}{2023} (\bibinfo{year}{1996}{\natexlab{a}}),
  \eprint{hep-ph/9512218}.

\bibitem[{\citenamefont{Vogelsang}(1996{\natexlab{b}})}]{Vogelsang:1996im}
\bibinfo{author}{\bibfnamefont{W.}~\bibnamefont{Vogelsang}},
  \bibinfo{journal}{Nucl. Phys. B} \textbf{\bibinfo{volume}{475}},
  \bibinfo{pages}{47} (\bibinfo{year}{1996}{\natexlab{b}}),
  \eprint{hep-ph/9603366}.

\bibitem[{\citenamefont{Moch et~al.}(2014)\citenamefont{Moch, Vermaseren, and
  Vogt}}]{Moch:2014sna}
\bibinfo{author}{\bibfnamefont{S.}~\bibnamefont{Moch}},
  \bibinfo{author}{\bibfnamefont{J.~A.~M.} \bibnamefont{Vermaseren}},
  \bibnamefont{and} \bibinfo{author}{\bibfnamefont{A.}~\bibnamefont{Vogt}},
  \bibinfo{journal}{Nucl. Phys. B} \textbf{\bibinfo{volume}{889}},
  \bibinfo{pages}{351} (\bibinfo{year}{2014}), \eprint{1409.5131}.

\bibitem[{\citenamefont{Bl\"umlein et~al.}(2021)\citenamefont{Bl\"umlein,
  Marquard, Schneider, and Sch\"onwald}}]{Blumlein:2021enk}
\bibinfo{author}{\bibfnamefont{J.}~\bibnamefont{Bl\"umlein}},
  \bibinfo{author}{\bibfnamefont{P.}~\bibnamefont{Marquard}},
  \bibinfo{author}{\bibfnamefont{C.}~\bibnamefont{Schneider}},
  \bibnamefont{and}
  \bibinfo{author}{\bibfnamefont{K.}~\bibnamefont{Sch\"onwald}},
  \bibinfo{journal}{Nucl. Phys. B} \textbf{\bibinfo{volume}{971}},
  \bibinfo{pages}{115542} (\bibinfo{year}{2021}), \eprint{2107.06267}.

\bibitem[{\citenamefont{Bl\"umlein
  et~al.}(2022{\natexlab{a}})\citenamefont{Bl\"umlein, Marquard, Schneider, and
  Sch\"onwald}}]{Blumlein:2021ryt}
\bibinfo{author}{\bibfnamefont{J.}~\bibnamefont{Bl\"umlein}},
  \bibinfo{author}{\bibfnamefont{P.}~\bibnamefont{Marquard}},
  \bibinfo{author}{\bibfnamefont{C.}~\bibnamefont{Schneider}},
  \bibnamefont{and}
  \bibinfo{author}{\bibfnamefont{K.}~\bibnamefont{Sch\"onwald}},
  \bibinfo{journal}{JHEP} \textbf{\bibinfo{volume}{01}}, \bibinfo{pages}{193}
  (\bibinfo{year}{2022}{\natexlab{a}}), \eprint{2111.12401}.

\bibitem[{\citenamefont{Bl\"umlein
  et~al.}(2022{\natexlab{b}})\citenamefont{Bl\"umlein, Marquard, Schneider, and
  Sch\"onwald}}]{Blumlein:2022gpp}
\bibinfo{author}{\bibfnamefont{J.}~\bibnamefont{Bl\"umlein}},
  \bibinfo{author}{\bibfnamefont{P.}~\bibnamefont{Marquard}},
  \bibinfo{author}{\bibfnamefont{C.}~\bibnamefont{Schneider}},
  \bibnamefont{and}
  \bibinfo{author}{\bibfnamefont{K.}~\bibnamefont{Sch\"onwald}},
  \bibinfo{journal}{JHEP} \textbf{\bibinfo{volume}{11}}, \bibinfo{pages}{156}
  (\bibinfo{year}{2022}{\natexlab{b}}), \eprint{2208.14325}.

\bibitem[{\citenamefont{Almasy et~al.}(2012)\citenamefont{Almasy, Moch, and
  Vogt}}]{Almasy:2011eq}
\bibinfo{author}{\bibfnamefont{A.~A.} \bibnamefont{Almasy}},
  \bibinfo{author}{\bibfnamefont{S.}~\bibnamefont{Moch}}, \bibnamefont{and}
  \bibinfo{author}{\bibfnamefont{A.}~\bibnamefont{Vogt}},
  \bibinfo{journal}{Nucl. Phys. B} \textbf{\bibinfo{volume}{854}},
  \bibinfo{pages}{133} (\bibinfo{year}{2012}), \eprint{1107.2263}.

\bibitem[{\citenamefont{Chen et~al.}(2021)\citenamefont{Chen, Yang, Zhu, and
  Zhu}}]{Chen:2020uvt}
\bibinfo{author}{\bibfnamefont{H.}~\bibnamefont{Chen}},
  \bibinfo{author}{\bibfnamefont{T.-Z.} \bibnamefont{Yang}},
  \bibinfo{author}{\bibfnamefont{H.~X.} \bibnamefont{Zhu}}, \bibnamefont{and}
  \bibinfo{author}{\bibfnamefont{Y.~J.} \bibnamefont{Zhu}},
  \bibinfo{journal}{Chin. Phys. C} \textbf{\bibinfo{volume}{45}},
  \bibinfo{pages}{043101} (\bibinfo{year}{2021}), \eprint{2006.10534}.

\bibitem[{\citenamefont{Matiounine et~al.}(1998)\citenamefont{Matiounine,
  Smith, and van Neerven}}]{Matiounine:1998re}
\bibinfo{author}{\bibfnamefont{Y.}~\bibnamefont{Matiounine}},
  \bibinfo{author}{\bibfnamefont{J.}~\bibnamefont{Smith}}, \bibnamefont{and}
  \bibinfo{author}{\bibfnamefont{W.~L.} \bibnamefont{van Neerven}},
  \bibinfo{journal}{Phys. Rev. D} \textbf{\bibinfo{volume}{58}},
  \bibinfo{pages}{076002} (\bibinfo{year}{1998}), \eprint{hep-ph/9803439}.

\bibitem[{\citenamefont{Ravindran et~al.}(2004)\citenamefont{Ravindran, Smith,
  and van Neerven}}]{Ravindran:2003gi}
\bibinfo{author}{\bibfnamefont{V.}~\bibnamefont{Ravindran}},
  \bibinfo{author}{\bibfnamefont{J.}~\bibnamefont{Smith}}, \bibnamefont{and}
  \bibinfo{author}{\bibfnamefont{W.~L.} \bibnamefont{van Neerven}},
  \bibinfo{journal}{Nucl. Phys. B} \textbf{\bibinfo{volume}{682}},
  \bibinfo{pages}{421} (\bibinfo{year}{2004}), \eprint{hep-ph/0311304}.

\bibitem[{\citenamefont{Goyal et~al.}(to appear)\citenamefont{Goyal, Lee, Moch,
  Pathak, Rana, and Ravindran}}]{Goyal:2025xxx}
\bibinfo{author}{\bibfnamefont{S.}~\bibnamefont{Goyal}},
  \bibinfo{author}{\bibfnamefont{R.~N.} \bibnamefont{Lee}},
  \bibinfo{author}{\bibfnamefont{S.-O.} \bibnamefont{Moch}},
  \bibinfo{author}{\bibfnamefont{V.}~\bibnamefont{Pathak}},
  \bibinfo{author}{\bibfnamefont{N.}~\bibnamefont{Rana}}, \bibnamefont{and}
  \bibinfo{author}{\bibfnamefont{V.}~\bibnamefont{Ravindran}}
  (\bibinfo{year}{to appear}).

\bibitem[{\citenamefont{Larin and Vermaseren}(1993)}]{Larin:1993tp}
\bibinfo{author}{\bibfnamefont{S.~A.} \bibnamefont{Larin}} \bibnamefont{and}
  \bibinfo{author}{\bibfnamefont{J.~A.~M.} \bibnamefont{Vermaseren}},
  \bibinfo{journal}{Phys. Lett. B} \textbf{\bibinfo{volume}{303}},
  \bibinfo{pages}{334} (\bibinfo{year}{1993}), \eprint{hep-ph/9302208}.

\bibitem[{\citenamefont{Ahmed et~al.}(2015)\citenamefont{Ahmed, Gehrmann,
  Mathews, Rana, and Ravindran}}]{Ahmed:2015qpa}
\bibinfo{author}{\bibfnamefont{T.}~\bibnamefont{Ahmed}},
  \bibinfo{author}{\bibfnamefont{T.}~\bibnamefont{Gehrmann}},
  \bibinfo{author}{\bibfnamefont{P.}~\bibnamefont{Mathews}},
  \bibinfo{author}{\bibfnamefont{N.}~\bibnamefont{Rana}}, \bibnamefont{and}
  \bibinfo{author}{\bibfnamefont{V.}~\bibnamefont{Ravindran}},
  \bibinfo{journal}{JHEP} \textbf{\bibinfo{volume}{11}}, \bibinfo{pages}{169}
  (\bibinfo{year}{2015}), \eprint{1510.01715}.

\bibitem[{\citenamefont{Lee et~al.}(2022)\citenamefont{Lee, von Manteuffel,
  Schabinger, Smirnov, Smirnov, and Steinhauser}}]{Lee:2022nhh}
\bibinfo{author}{\bibfnamefont{R.~N.} \bibnamefont{Lee}},
  \bibinfo{author}{\bibfnamefont{A.}~\bibnamefont{von Manteuffel}},
  \bibinfo{author}{\bibfnamefont{R.~M.} \bibnamefont{Schabinger}},
  \bibinfo{author}{\bibfnamefont{A.~V.} \bibnamefont{Smirnov}},
  \bibinfo{author}{\bibfnamefont{V.~A.} \bibnamefont{Smirnov}},
  \bibnamefont{and}
  \bibinfo{author}{\bibfnamefont{M.}~\bibnamefont{Steinhauser}},
  \bibinfo{journal}{Phys. Rev. Lett.} \textbf{\bibinfo{volume}{128}},
  \bibinfo{pages}{212002} (\bibinfo{year}{2022}), \eprint{2202.04660}.

\bibitem[{\citenamefont{Nogueira}(1993)}]{Nogueira:1991ex}
\bibinfo{author}{\bibfnamefont{P.}~\bibnamefont{Nogueira}},
  \bibinfo{journal}{J. Comput. Phys.} \textbf{\bibinfo{volume}{105}},
  \bibinfo{pages}{279} (\bibinfo{year}{1993}).

\bibitem[{\citenamefont{Kuipers et~al.}(2013)\citenamefont{Kuipers, Ueda,
  Vermaseren, and Vollinga}}]{Kuipers:2012rf}
\bibinfo{author}{\bibfnamefont{J.}~\bibnamefont{Kuipers}},
  \bibinfo{author}{\bibfnamefont{T.}~\bibnamefont{Ueda}},
  \bibinfo{author}{\bibfnamefont{J.~A.~M.} \bibnamefont{Vermaseren}},
  \bibnamefont{and} \bibinfo{author}{\bibfnamefont{J.}~\bibnamefont{Vollinga}},
  \bibinfo{journal}{Comput. Phys. Commun.} \textbf{\bibinfo{volume}{184}},
  \bibinfo{pages}{1453} (\bibinfo{year}{2013}), \eprint{1203.6543}.

\bibitem[{\citenamefont{Ruijl et~al.}(2017)\citenamefont{Ruijl, Ueda, and
  Vermaseren}}]{Ruijl:2017dtg}
\bibinfo{author}{\bibfnamefont{B.}~\bibnamefont{Ruijl}},
  \bibinfo{author}{\bibfnamefont{T.}~\bibnamefont{Ueda}}, \bibnamefont{and}
  \bibinfo{author}{\bibfnamefont{J.}~\bibnamefont{Vermaseren}}
  (\bibinfo{year}{2017}), \eprint{1707.06453}.

\bibitem[{\citenamefont{Anastasiou et~al.}(2004)\citenamefont{Anastasiou,
  Melnikov, and Petriello}}]{Anastasiou:2003gr}
\bibinfo{author}{\bibfnamefont{C.}~\bibnamefont{Anastasiou}},
  \bibinfo{author}{\bibfnamefont{K.}~\bibnamefont{Melnikov}}, \bibnamefont{and}
  \bibinfo{author}{\bibfnamefont{F.}~\bibnamefont{Petriello}},
  \bibinfo{journal}{Phys. Rev. D} \textbf{\bibinfo{volume}{69}},
  \bibinfo{pages}{076010} (\bibinfo{year}{2004}), \eprint{hep-ph/0311311}.

\bibitem[{\citenamefont{Anastasiou et~al.}(2012)\citenamefont{Anastasiou,
  Buehler, Duhr, and Herzog}}]{Anastasiou:2012kq}
\bibinfo{author}{\bibfnamefont{C.}~\bibnamefont{Anastasiou}},
  \bibinfo{author}{\bibfnamefont{S.}~\bibnamefont{Buehler}},
  \bibinfo{author}{\bibfnamefont{C.}~\bibnamefont{Duhr}}, \bibnamefont{and}
  \bibinfo{author}{\bibfnamefont{F.}~\bibnamefont{Herzog}},
  \bibinfo{journal}{JHEP} \textbf{\bibinfo{volume}{11}}, \bibinfo{pages}{062}
  (\bibinfo{year}{2012}), \eprint{1208.3130}.

\bibitem[{\citenamefont{Chetyrkin and Tkachov}(1981)}]{Chetyrkin:1981qh}
\bibinfo{author}{\bibfnamefont{K.~G.} \bibnamefont{Chetyrkin}}
  \bibnamefont{and} \bibinfo{author}{\bibfnamefont{F.~V.}
  \bibnamefont{Tkachov}}, \bibinfo{journal}{Nucl. Phys. B}
  \textbf{\bibinfo{volume}{192}}, \bibinfo{pages}{159} (\bibinfo{year}{1981}).

\bibitem[{\citenamefont{Laporta}(2000)}]{Laporta:2001dd}
\bibinfo{author}{\bibfnamefont{S.}~\bibnamefont{Laporta}},
  \bibinfo{journal}{Int. J. Mod. Phys. A} \textbf{\bibinfo{volume}{15}},
  \bibinfo{pages}{5087} (\bibinfo{year}{2000}), \eprint{hep-ph/0102033}.

\bibitem[{\citenamefont{Lee}(2014)}]{Lee:2013mka}
\bibinfo{author}{\bibfnamefont{R.~N.} \bibnamefont{Lee}}, \bibinfo{journal}{J.
  Phys. Conf. Ser.} \textbf{\bibinfo{volume}{523}}, \bibinfo{pages}{012059}
  (\bibinfo{year}{2014}), \eprint{1310.1145}.

\bibitem[{\citenamefont{Matsuura et~al.}(1989)\citenamefont{Matsuura, van~der
  Marck, and van Neerven}}]{Matsuura:1988sm}
\bibinfo{author}{\bibfnamefont{T.}~\bibnamefont{Matsuura}},
  \bibinfo{author}{\bibfnamefont{S.~C.} \bibnamefont{van~der Marck}},
  \bibnamefont{and} \bibinfo{author}{\bibfnamefont{W.~L.} \bibnamefont{van
  Neerven}}, \bibinfo{journal}{Nucl. Phys. B} \textbf{\bibinfo{volume}{319}},
  \bibinfo{pages}{570} (\bibinfo{year}{1989}).

\bibitem[{\citenamefont{Zijlstra and van Neerven}(1992)}]{Zijlstra:1992qd}
\bibinfo{author}{\bibfnamefont{E.~B.} \bibnamefont{Zijlstra}} \bibnamefont{and}
  \bibinfo{author}{\bibfnamefont{W.~L.} \bibnamefont{van Neerven}},
  \bibinfo{journal}{Nucl. Phys. B} \textbf{\bibinfo{volume}{383}},
  \bibinfo{pages}{525} (\bibinfo{year}{1992}).

\bibitem[{\citenamefont{Rijken and van Neerven}(1997)}]{Rijken:1996ns}
\bibinfo{author}{\bibfnamefont{P.~J.} \bibnamefont{Rijken}} \bibnamefont{and}
  \bibinfo{author}{\bibfnamefont{W.~L.} \bibnamefont{van Neerven}},
  \bibinfo{journal}{Nucl. Phys. B} \textbf{\bibinfo{volume}{487}},
  \bibinfo{pages}{233} (\bibinfo{year}{1997}), \eprint{hep-ph/9609377}.

\bibitem[{\citenamefont{Ravindran et~al.}(2003)\citenamefont{Ravindran, Smith,
  and van Neerven}}]{Ravindran:2003um}
\bibinfo{author}{\bibfnamefont{V.}~\bibnamefont{Ravindran}},
  \bibinfo{author}{\bibfnamefont{J.}~\bibnamefont{Smith}}, \bibnamefont{and}
  \bibinfo{author}{\bibfnamefont{W.~L.} \bibnamefont{van Neerven}},
  \bibinfo{journal}{Nucl. Phys. B} \textbf{\bibinfo{volume}{665}},
  \bibinfo{pages}{325} (\bibinfo{year}{2003}), \eprint{hep-ph/0302135}.

\bibitem[{\citenamefont{Kotikov}(1991)}]{Kotikov:1990kg}
\bibinfo{author}{\bibfnamefont{A.~V.} \bibnamefont{Kotikov}},
  \bibinfo{journal}{Phys. Lett. B} \textbf{\bibinfo{volume}{254}},
  \bibinfo{pages}{158} (\bibinfo{year}{1991}).

\bibitem[{\citenamefont{Argeri and Mastrolia}(2007)}]{Argeri:2007up}
\bibinfo{author}{\bibfnamefont{M.}~\bibnamefont{Argeri}} \bibnamefont{and}
  \bibinfo{author}{\bibfnamefont{P.}~\bibnamefont{Mastrolia}},
  \bibinfo{journal}{Int. J. Mod. Phys. A} \textbf{\bibinfo{volume}{22}},
  \bibinfo{pages}{4375} (\bibinfo{year}{2007}), \eprint{0707.4037}.

\bibitem[{\citenamefont{Remiddi}(1997)}]{Remiddi:1997ny}
\bibinfo{author}{\bibfnamefont{E.}~\bibnamefont{Remiddi}},
  \bibinfo{journal}{Nuovo Cim. A} \textbf{\bibinfo{volume}{110}},
  \bibinfo{pages}{1435} (\bibinfo{year}{1997}), \eprint{hep-th/9711188}.

\bibitem[{\citenamefont{Henn}(2013)}]{Henn:2013pwa}
\bibinfo{author}{\bibfnamefont{J.~M.} \bibnamefont{Henn}},
  \bibinfo{journal}{Phys. Rev. Lett.} \textbf{\bibinfo{volume}{110}},
  \bibinfo{pages}{251601} (\bibinfo{year}{2013}), \eprint{1304.1806}.

\bibitem[{\citenamefont{Ablinger et~al.}(2016)\citenamefont{Ablinger, Behring,
  Bl\"umlein, De~Freitas, von Manteuffel, and Schneider}}]{Ablinger:2015tua}
\bibinfo{author}{\bibfnamefont{J.}~\bibnamefont{Ablinger}},
  \bibinfo{author}{\bibfnamefont{A.}~\bibnamefont{Behring}},
  \bibinfo{author}{\bibfnamefont{J.}~\bibnamefont{Bl\"umlein}},
  \bibinfo{author}{\bibfnamefont{A.}~\bibnamefont{De~Freitas}},
  \bibinfo{author}{\bibfnamefont{A.}~\bibnamefont{von Manteuffel}},
  \bibnamefont{and}
  \bibinfo{author}{\bibfnamefont{C.}~\bibnamefont{Schneider}},
  \bibinfo{journal}{Comput. Phys. Commun.} \textbf{\bibinfo{volume}{202}},
  \bibinfo{pages}{33} (\bibinfo{year}{2016}), \eprint{1509.08324}.

\bibitem[{\citenamefont{Lee}(2021)}]{Lee:2020zfb}
\bibinfo{author}{\bibfnamefont{R.~N.} \bibnamefont{Lee}},
  \bibinfo{journal}{Comput. Phys. Commun.} \textbf{\bibinfo{volume}{267}},
  \bibinfo{pages}{108058} (\bibinfo{year}{2021}), \eprint{2012.00279}.

\bibitem[{\citenamefont{Duplancic and Nizic}(2001)}]{Duplancic:2000sk}
\bibinfo{author}{\bibfnamefont{G.}~\bibnamefont{Duplancic}} \bibnamefont{and}
  \bibinfo{author}{\bibfnamefont{B.}~\bibnamefont{Nizic}},
  \bibinfo{journal}{Eur. Phys. J. C} \textbf{\bibinfo{volume}{20}},
  \bibinfo{pages}{357} (\bibinfo{year}{2001}), \eprint{hep-ph/0006249}.

\bibitem[{\citenamefont{Gehrmann and Remiddi}(2002)}]{Gehrmann:2002zr}
\bibinfo{author}{\bibfnamefont{T.}~\bibnamefont{Gehrmann}} \bibnamefont{and}
  \bibinfo{author}{\bibfnamefont{E.}~\bibnamefont{Remiddi}},
  \bibinfo{journal}{Nucl. Phys. B} \textbf{\bibinfo{volume}{640}},
  \bibinfo{pages}{379} (\bibinfo{year}{2002}), \eprint{hep-ph/0207020}.

\bibitem[{\citenamefont{Gehrmann and Sch\"urmann}(2022)}]{Gehrmann:2022cih}
\bibinfo{author}{\bibfnamefont{T.}~\bibnamefont{Gehrmann}} \bibnamefont{and}
  \bibinfo{author}{\bibfnamefont{R.}~\bibnamefont{Sch\"urmann}},
  \bibinfo{journal}{JHEP} \textbf{\bibinfo{volume}{04}}, \bibinfo{pages}{031}
  (\bibinfo{year}{2022}), \eprint{2201.06982}.

\bibitem[{\citenamefont{Ravindran et~al.}(2007)\citenamefont{Ravindran, Smith,
  and van Neerven}}]{Ravindran:2006bu}
\bibinfo{author}{\bibfnamefont{V.}~\bibnamefont{Ravindran}},
  \bibinfo{author}{\bibfnamefont{J.}~\bibnamefont{Smith}}, \bibnamefont{and}
  \bibinfo{author}{\bibfnamefont{W.~L.} \bibnamefont{van Neerven}},
  \bibinfo{journal}{Nucl. Phys. B} \textbf{\bibinfo{volume}{767}},
  \bibinfo{pages}{100} (\bibinfo{year}{2007}), \eprint{hep-ph/0608308}.

\bibitem[{\citenamefont{Ahmed et~al.}(2014)\citenamefont{Ahmed, Mandal, Rana,
  and Ravindran}}]{Ahmed:2014uya}
\bibinfo{author}{\bibfnamefont{T.}~\bibnamefont{Ahmed}},
  \bibinfo{author}{\bibfnamefont{M.~K.} \bibnamefont{Mandal}},
  \bibinfo{author}{\bibfnamefont{N.}~\bibnamefont{Rana}}, \bibnamefont{and}
  \bibinfo{author}{\bibfnamefont{V.}~\bibnamefont{Ravindran}},
  \bibinfo{journal}{Phys. Rev. Lett.} \textbf{\bibinfo{volume}{113}},
  \bibinfo{pages}{212003} (\bibinfo{year}{2014}), \eprint{1404.6504}.

\bibitem[{\citenamefont{A~H et~al.}(2021)\citenamefont{A~H, Mukherjee,
  Ravindran, Sankar, and Tiwari}}]{AH:2020qoa}
\bibinfo{author}{\bibfnamefont{A.}~\bibnamefont{A~H}},
  \bibinfo{author}{\bibfnamefont{P.}~\bibnamefont{Mukherjee}},
  \bibinfo{author}{\bibfnamefont{V.}~\bibnamefont{Ravindran}},
  \bibinfo{author}{\bibfnamefont{A.}~\bibnamefont{Sankar}}, \bibnamefont{and}
  \bibinfo{author}{\bibfnamefont{S.}~\bibnamefont{Tiwari}},
  \bibinfo{journal}{Phys. Rev. D} \textbf{\bibinfo{volume}{103}},
  \bibinfo{pages}{L111502} (\bibinfo{year}{2021}), \eprint{2010.00079}.

\bibitem[{\citenamefont{Bertone et~al.}(2024)\citenamefont{Bertone, Chiefa, and
  Nocera}}]{Bertone:2024taw}
\bibinfo{author}{\bibfnamefont{V.}~\bibnamefont{Bertone}},
  \bibinfo{author}{\bibfnamefont{A.}~\bibnamefont{Chiefa}}, \bibnamefont{and}
  \bibinfo{author}{\bibfnamefont{E.~R.} \bibnamefont{Nocera}}
  (\bibinfo{collaboration}{MAP}) (\bibinfo{year}{2024}), \eprint{2404.04712}.

\bibitem[{\citenamefont{Bertone et~al.}(2017)\citenamefont{Bertone, Carrazza,
  Hartland, Nocera, and Rojo}}]{Bertone:2017tyb}
\bibinfo{author}{\bibfnamefont{V.}~\bibnamefont{Bertone}},
  \bibinfo{author}{\bibfnamefont{S.}~\bibnamefont{Carrazza}},
  \bibinfo{author}{\bibfnamefont{N.~P.} \bibnamefont{Hartland}},
  \bibinfo{author}{\bibfnamefont{E.~R.} \bibnamefont{Nocera}},
  \bibnamefont{and} \bibinfo{author}{\bibfnamefont{J.}~\bibnamefont{Rojo}}
  (\bibinfo{collaboration}{NNPDF}), \bibinfo{journal}{Eur. Phys. J. C}
  \textbf{\bibinfo{volume}{77}}, \bibinfo{pages}{516} (\bibinfo{year}{2017}),
  \eprint{1706.07049}.

\bibitem[{\citenamefont{Abdul~Khalek
  et~al.}(2022{\natexlab{b}})\citenamefont{Abdul~Khalek, Bertone, Khoudli, and
  Nocera}}]{AbdulKhalek:2022laj}
\bibinfo{author}{\bibfnamefont{R.}~\bibnamefont{Abdul~Khalek}},
  \bibinfo{author}{\bibfnamefont{V.}~\bibnamefont{Bertone}},
  \bibinfo{author}{\bibfnamefont{A.}~\bibnamefont{Khoudli}}, \bibnamefont{and}
  \bibinfo{author}{\bibfnamefont{E.~R.} \bibnamefont{Nocera}}
  (\bibinfo{collaboration}{MAP (Multi-dimensional Analyses of Partonic
  distributions)}), \bibinfo{journal}{Phys. Lett. B}
  \textbf{\bibinfo{volume}{834}}, \bibinfo{pages}{137456}
  (\bibinfo{year}{2022}{\natexlab{b}}), \eprint{2204.10331}.

\bibitem[{\citenamefont{Ball et~al.}(2017)}]{NNPDF:2017mvq}
\bibinfo{author}{\bibfnamefont{R.~D.} \bibnamefont{Ball}} \bibnamefont{et~al.}
  (\bibinfo{collaboration}{NNPDF}), \bibinfo{journal}{Eur. Phys. J. C}
  \textbf{\bibinfo{volume}{77}}, \bibinfo{pages}{663} (\bibinfo{year}{2017}),
  \eprint{1706.00428}.

\bibitem[{\citenamefont{Alekseev et~al.}(2010)}]{COMPASS:2010hwr}
\bibinfo{author}{\bibfnamefont{M.~G.} \bibnamefont{Alekseev}}
  \bibnamefont{et~al.} (\bibinfo{collaboration}{COMPASS}),
  \bibinfo{journal}{Phys. Lett. B} \textbf{\bibinfo{volume}{693}},
  \bibinfo{pages}{227} (\bibinfo{year}{2010}), \eprint{1007.4061}.

\bibitem[{\citenamefont{Bonino et~al.}(2024{\natexlab{b}})\citenamefont{Bonino,
  Gehrmann, L\"ochner, Sch\"onwald, and Stagnitto}}]{Bonino:2024wgg}
\bibinfo{author}{\bibfnamefont{L.}~\bibnamefont{Bonino}},
  \bibinfo{author}{\bibfnamefont{T.}~\bibnamefont{Gehrmann}},
  \bibinfo{author}{\bibfnamefont{M.}~\bibnamefont{L\"ochner}},
  \bibinfo{author}{\bibfnamefont{K.}~\bibnamefont{Sch\"onwald}},
  \bibnamefont{and}
  \bibinfo{author}{\bibfnamefont{G.}~\bibnamefont{Stagnitto}},
  \bibinfo{journal}{Phys. Rev. Lett.} \textbf{\bibinfo{volume}{133}},
  \bibinfo{pages}{211904} (\bibinfo{year}{2024}{\natexlab{b}}),
  \eprint{2404.08597}.

\bibitem[{\citenamefont{Bonino et~al.}(2025)\citenamefont{Bonino,
  Gehrmann, L\"ochner, Sch\"onwald, and Stagnitto}}]{Bonino:2025bqa}
\bibinfo{author}{\bibfnamefont{L.}~\bibnamefont{Bonino}},
  \bibinfo{author}{\bibfnamefont{T.}~\bibnamefont{Gehrmann}},
  \bibinfo{author}{\bibfnamefont{M.}~\bibnamefont{L\"ochner}},
  \bibinfo{author}{\bibfnamefont{K.}~\bibnamefont{Sch\"onwald}},
  \bibnamefont{and}
  \bibinfo{author}{\bibfnamefont{G.}~\bibnamefont{Stagnitto}}
  (\bibinfo{year}{2025}),
  \eprint{2510.00100}.
  
\end{thebibliography}

\end{document}